\documentclass{elsart}

\input{tcilatex}
\begin{document}

\begin{frontmatter}
\title{Tail universalities in rank distributions as an algebraic problem:
the beta-like function}
\author{G.G. Naumis}
\ead{naumis@fisica.unam.mx}
\address{Departamento de Fisica-Quimica, Instituto de Fisica. 
Universidad Nacional Aut\'{o}noma de M\'{e}xico. Apdo.
Postal 20-364, 01000, M\'{e}xico D.F., Mexico.}
\author{G. Cocho}
\address{Departamento de Sistemas Complejos, Instituto de Fisica. 
Universidad Nacional Aut\'{o}noma de M\'{e}xico. Apdo.
Postal 20-364, 01000, M\'{e}xico D.F., Mexico.}
\date{\today }

\begin{abstract}
Although power laws of the Zipf type have been used by many workers to fit rank distributions in different fields 
like in economy, geophysics, genetics, soft-matter, networks etc., these fits usually fail at 
the tails. Some distributions have been proposed to solve the problem, but
unfortunately they do not fit at the same time both ending tails.
We show that many different data in rank laws, like in granular
materials, codons, author impact in scientific journal, etc. are very well
fitted by a beta-like function. Then we propose that such universality
is due to the fact that a system made from many subsystems or choices,
imply stretched exponential frequency-rank functions
which qualitatively and quantitatively can be fitted with the proposed beta-like function
distribution in the limit of many random variables. We prove this by
transforming the problem into an algebraic one: finding the rank of
successive products of a given set of numbers.
\end{abstract}


\end{frontmatter}

\section{Introduction}

Both natural language texts and coding DNA sequences present power laws in
the observed frequency of a word as a function of its rank ($r$), where the
rank is just the ordinal position of a word if all words are ordered
according to their decreasing frequency. Usually, the most frequent word has
rank $1$, the next most frequent rank $2$ and so on. This power law behavior
of the ranking is known as the Zipf law \cite{Li}, and it is very common in
physics, biology, geography, economics, linguistics, etc. \cite{Li}. In
physics one can cite the rank distribution of stick-slip events in sheared
granular media \cite{Bretz}, earthquakes \cite{Bretz}, radionuclides
half-life time and nuclides mass number \cite{Nuclear}. Many complex systems
share as well the same phenomenology, as happens in networks \cite{Fortunato}%
, biological clocks \cite{Yang} and metabolic networks \cite{Jeong}. Zipf
discovered his rank law by analyzing manually the frequencies of 29,899
different words types in the novel \textquotedblright
Ulysses\textquotedblright\ by James Joyce, but when a larger set of words is
considered, a deviation from a power law is observed for larger ranks \cite%
{Lequan}. A similar behavior is found in coding genetic sequences.
Deviations from the Zipf law are also found in the tails ranking of many
physical systems \cite{Sornette}. In fact, is clear that one should expect a
different behavior at the tails, since finite size effects should be present
and the power law must be "stopped" at a certain region. In spite of this,
many workers just ignore the tail effects by fitting the data in a \
restricted range, or they proceed in a very questionable way by fitting all
the data with a power law. Others have fitted sets of data in nature and in
economy with stretched exponentials \cite{Sornette} and log-normal
distributions \cite{Montroll}. \ The problem with the previous expressions
is that they do not fit the data at \textit{both ending tails}, where
different kinds of processes are set in once a crossover region is reached.
Such crossovers are due to finite size effects, in which different
mechanisms are set in when certain big and small scales are reached. This
leads to the idea of using multiscaling physical modelling to understand
such features. Maybe the best example of the previous situation occurs in
turbulence, where Kolmogorov%
\'{}%
s power law is observed only in the inertial regimen \cite{Kolmogorov1}\cite%
{Kolmogorov2}. In one tail (small length scales) energy dissipation plays
the main role, while energy injection dominates at big scales \cite%
{Kolmogorov1}\cite{Kolmogorov2}. For each of these limits, the scaling
behavior is different \cite{Kolmogorov3}\cite{Warhaft}. One can conjecture
that similar ideas are behind many other complex physical systems, since we
report that many rank laws are extremely well parametrized, outperforming
many other rank-order models, with a two exponent beta function-like formula
with parameters $\{a,b\}$,%
\begin{equation}
f(r)=K\frac{(R-r+1)^{b}}{r^{a}},  \label{beta}
\end{equation}%
where $a$ and $b$ are fitted from the data, $r$ is the rank and $R$ is the
maximal $r$. If $f(r)$ is normalized to $1$, then $K\equiv 1/\sum_{r=1}^{N}$ 
$(R-r+1)^{b}/r^{a}$. For $R\gg 1$, $K$ can be transformed into an integral
that yields $K\approx \Gamma (b-a+2)/\Gamma (1-a)\Gamma (1+b)$. We will show
that $f(r)$ is related with a kind of central limit theorem, in which $a$
and $b$ seem to be parameters related with the onset of different
mechanisms. Our work is in the same spirit of Moyano \textit{et. al.} \cite%
{Moyano}, who have commented that the rather ubiquitous presence of the
Tsallis $q$-distributions is maybe due to a $q$-generalized central limit
theorem for a class of non independent, correlated, product of probability
distributions \cite{Marsh}. The outline of this paper is the following: in
section II we present some representative examples of the phenomenology that
we have observed. In section III we show how this phenomenology can be
studied as a problem of hierarchies in the product of random variables, and
then transformed into a related algebraic problem: what is the rank of a set
of numbers produced by the iterative product of an initial finite set of
numbers. In section IV, we solve the proposed problem, and finally, in
section V we give the conclusions of this work.

\section{Phenomenology of rank laws and the beta-like function}

As starting point, we will provide some representative results of the wide
phenomenology found in the tails of rank laws. We start with an example from
geography. Fig. 1 shows the population ranking of four representative
municipalities in Mexico and Spain in a semilog plot. The corresponding fits
using Eq. (\ref{beta}) are given by solid lines. The agreement is excellent,
with a correlation coefficient $R$ bigger than $0.98$ for all fits. The
values of $a$ and $b$ for each fit are shown in the inset of the plot. We
have verified that similar good results are obtained for the population of
countries and states.

\FRAME{ftbpFU}{3.6461in}{2.5598in}{0pt}{\Qcb{Population ranking of four
representative municipalites from Mexico and Spain. The solid lines are the
fits obtained from Eq. (\protect\ref{beta}) The inset presents the
corresponding values of $a$ and $b$ used in the fits.}}{}{population.eps}{%
\special{language "Scientific Word";type "GRAPHIC";maintain-aspect-ratio
TRUE;display "USEDEF";valid_file "F";width 3.6461in;height 2.5598in;depth
0pt;original-width 8.489in;original-height 5.9525in;cropleft "0";croptop
"1";cropright "1";cropbottom "0";filename '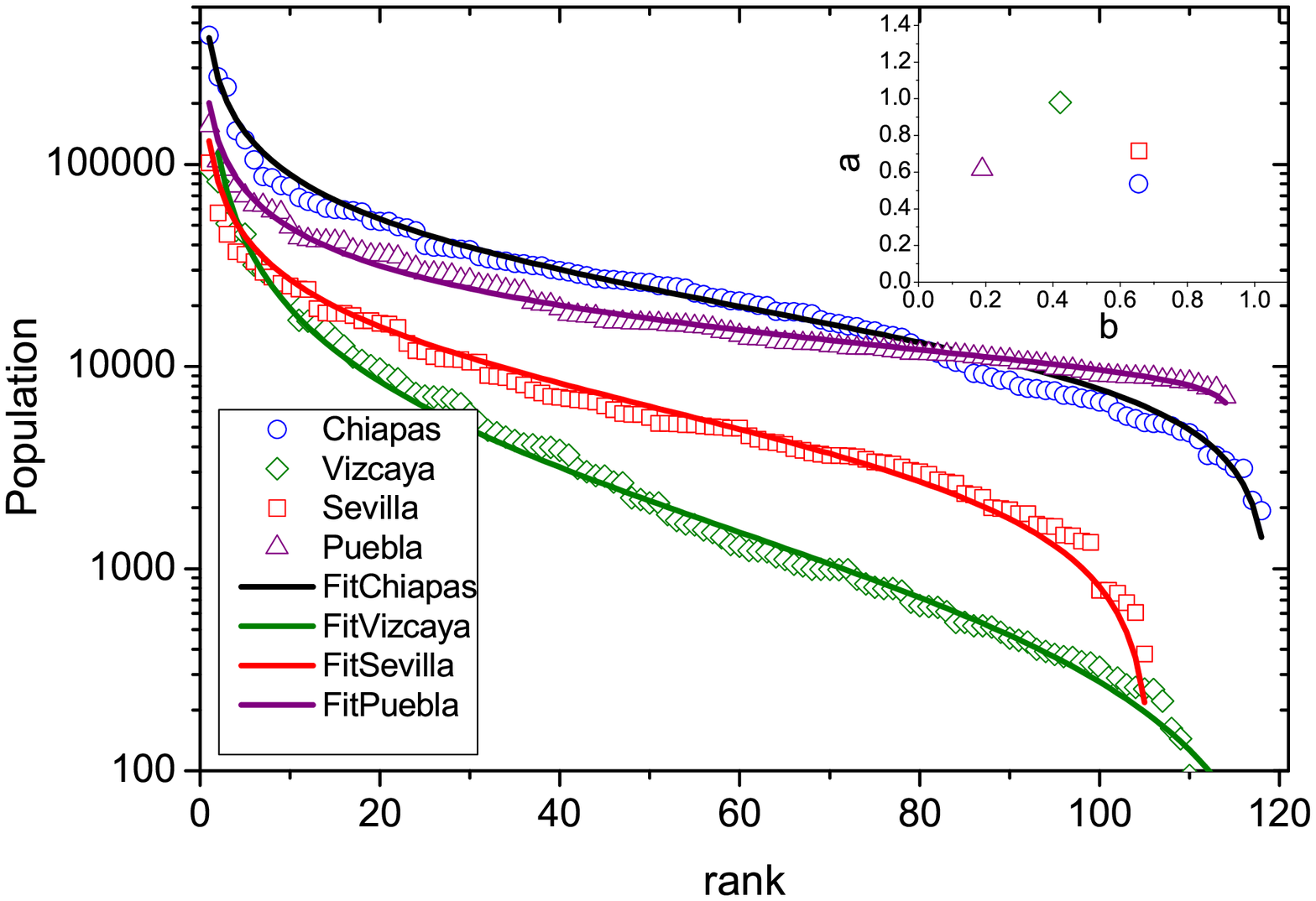';file-properties
"XNPEU";}}Figure 2 shows the impact factor against the rank of scientific
journals, taken from a recent study \cite{Pospescu}, compared with the fits
given by Eq. (\ref{beta}). Again, all the fits are excellent, with
correlation coefficients above $0.98$.\FRAME{ftbpFU}{3.2837in}{2.3073in}{0pt%
}{\Qcb{Impact factor as a function of the rank for physics, computer science
and agroscience. Fits using the beta-like function are shown as solid lines.
Inset: values of $a$ and $b$. }}{}{impact.eps}{\special{language "Scientific
Word";type "GRAPHIC";maintain-aspect-ratio TRUE;display "USEDEF";valid_file
"F";width 3.2837in;height 2.3073in;depth 0pt;original-width
8.489in;original-height 5.9525in;cropleft "0";croptop "1";cropright
"1";cropbottom "0";filename
'../germinal/Submision/impact.eps';file-properties "XNPEU";}}

Similar excellent fitting results are obtained for codon usage in genomes,
as shown in Fig. 3, where we plot the logarithm of the frequency of codons
(normalized to $1000$) as a function of the rank for different
representative organisms, taken from a genome database \cite{database}. For
all the organisms, the resulting correlation parameters are bigger than $%
0.97 $.

\bigskip

\FRAME{ftbpFU}{3.2214in}{2.2632in}{0pt}{\Qcb{Frequency of codons (normalized
to 1000) as a function of the rank for the genome of four different species,
with their correponding fits shown as solid lines. Inset: values of a $a$
and $b$ used for the fits in the beta-like distribution.}}{}{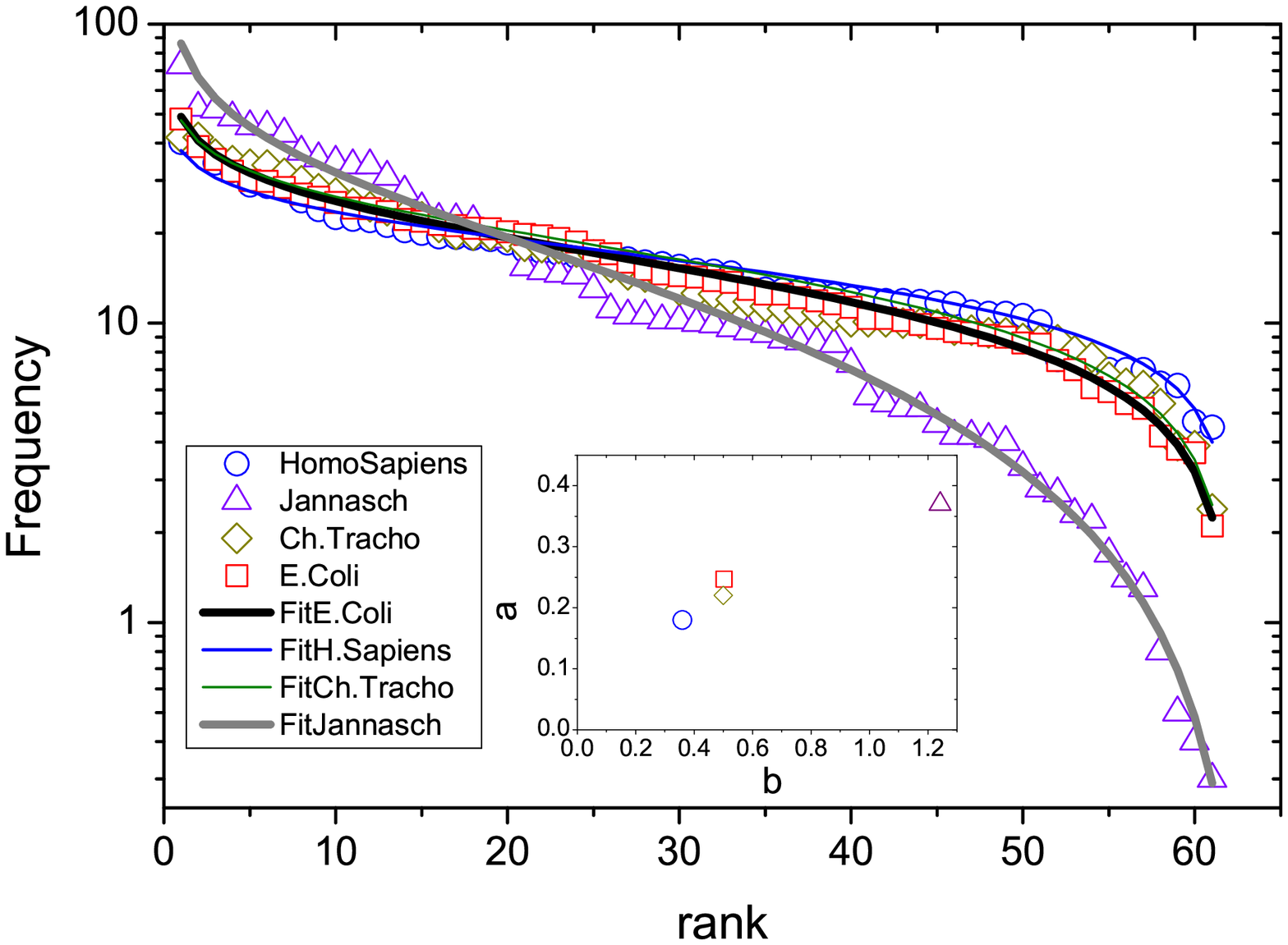}{%
\special{ language "Scientific Word"; type "GRAPHIC"; maintain-aspect-ratio
TRUE; display "USEDEF"; valid_file "F"; width 3.2214in; height 2.2632in;
depth 0pt; original-width 8.489in; original-height 5.9525in; cropleft "0";
croptop "1"; cropright "1"; cropbottom "0"; filename
'../germinal/Submision/codons.eps';file-properties "XNPEU";}}

Now we turn our attention to physics. In Fig. 4 we plot the rank-ordered
distribution of stick-slip events in a slowly sheared granular media taken
from Ref. \cite{Bretz}, fitted using\ Eq. (\ref{beta}). Although a modified
power law was proposed in Ref. \cite{Bretz} to explain the results, the
present fit also gives a better correlation coefficient.

\FRAME{ftbpFU}{3.4143in}{2.3981in}{0pt}{\Qcb{Rank-ordered distribution of
stick-slip events in a slowly sheared granular media. Circles are data taken
from Ref. \protect\cite{Bretz}, and the solid line is a fit using Eq. (%
\protect\ref{beta}), with $a=1.08$ and $b=0.40$}}{}{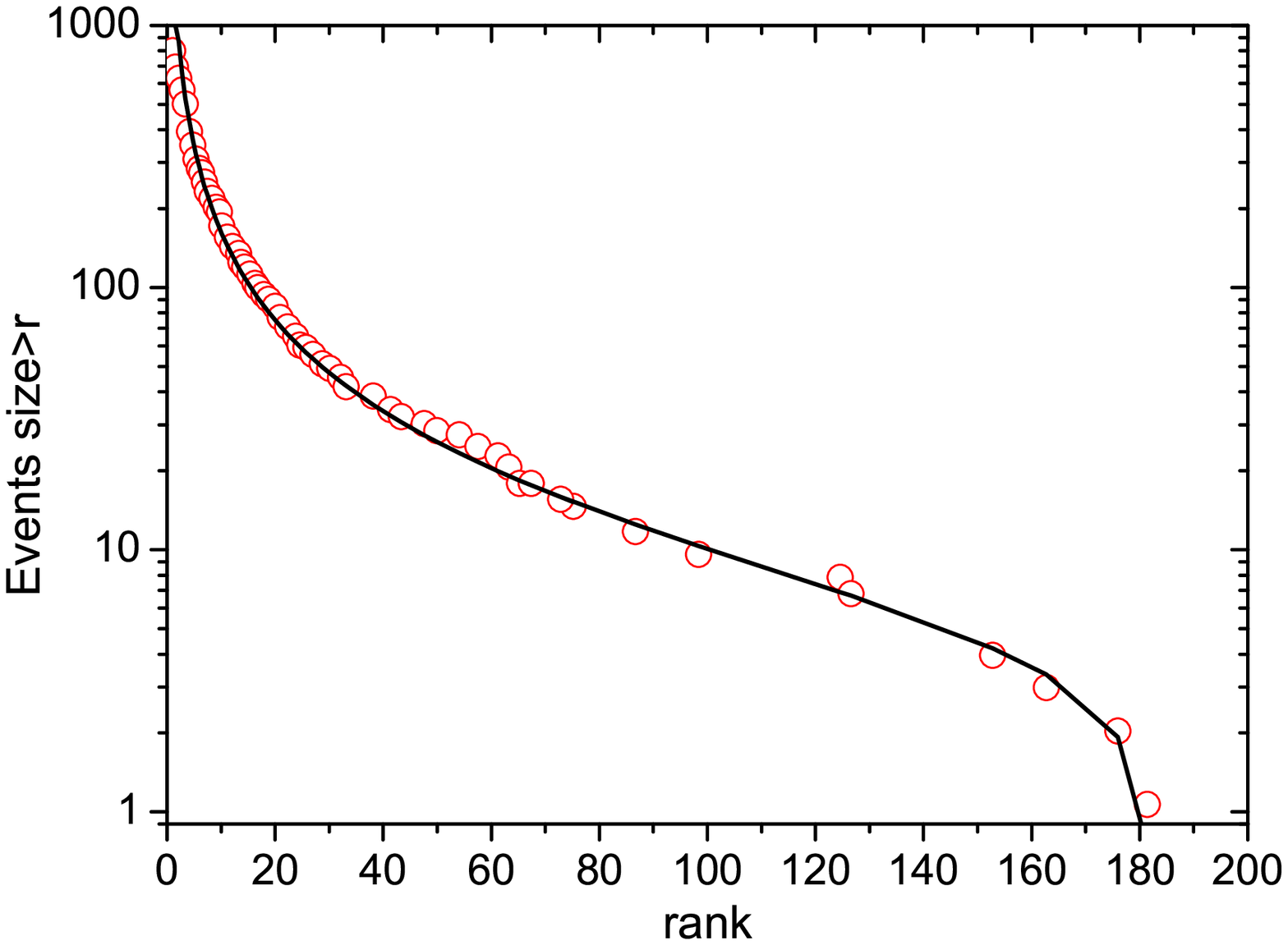}{\special%
{language "Scientific Word";type "GRAPHIC";maintain-aspect-ratio
TRUE;display "USEDEF";valid_file "F";width 3.4143in;height 2.3981in;depth
0pt;original-width 8.489in;original-height 5.9525in;cropleft "0";croptop
"1";cropright "1";cropbottom "0";filename
'../germinal/Submision/granular.eps';file-properties "XNPEU";}}

Here we presented four examples, but Eq. (\ref{beta}) can be used with
excellent results in order to correct the Gutenberg-Ritcher law in
earthquakes ranking, B\'{e}nard convection cells and in many different
fields, like arquitecture, music or roads \cite{Cocho}.

\section{\protect\bigskip Hierarchy in a multiplicative stochastic processes}

The previous section leads to the conclusion that both ending tails of the
ranking present some degree of universality, and Eq. (\ref{beta}) seems to
be an excellent fitting function due to the fact that it gives the right
shape of the curve and thus very good correlation coefficients. Also, it is
simple and can be reduced to a pure power law by using an appropriate choice
of $a$ and $b$. \ As the $\{a,b\}$ distributions is indeed ubiquitous, one
can try to associate it to some generic mechanism, as happens in the central
limit theorem or in the product of correlated probability distributions \cite%
{Moyano}. \ 

In the dynamics of population, scientific journal impact factor, codon usage
and stick-slip events, there are many important issues that determine the
behavior. In the case of the impact factor we can cite for example the
ability to select a good problem for investigation, the gift for writing
clear papers, etc. Similar comments would be valid for the dynamics of
granular media, as well as in economy, linguistics, genetics, etc. All of
the previous systems share a common feature: their complex nature, \textit{%
i.e.}, they are build from many subsystems or path choices that produce a
final result. One can try to model such complexity as follows. Consider a
system made from $N$ identical subsystems, where each can have $s$ different
states or choices with probability $p_{j}$, and $j=1,...,s.$ \ When $N$ such
subsystems are put together, the state space consists of all $s^{N\text{ \ }%
} $possible sequences of length $N$. \ If we do not care about the order of
the choices or states in the string, there are just $(N+s-1)!/s!(N-1)!$ 
\textit{different combinations}. For example, if a system is made from $N=2$
subsystems, where each has two states or choices, say $1$ or $0$, the
possible global states are $(0,0)$, $(1,0)$, $(0,1)$ and $(1,1)$, while
there are only three combinations: $(0,0)$, $(1,1)$ and $(1,0)$, the last
one has multiplicity $2$. Each combination has a certain probability that we
call \textit{reduced probabilities} $x_{N}(n_{1},n_{2},...,n_{s})$. The
multiplicity of \ each different state is given by the multinomial
coefficient $N!/(n_{1}!n_{2}!n_{3}!...n_{s}!)$, where $n_{j}$ is the number
of subsystems in the $j$-esim state. The probability of a global state of
the whole system is,%
\begin{equation}
P_{N}(n_{1},n_{2},...,n_{s})=\frac{N!}{n_{1}!n_{2}!n_{3}!...n_{s}!}%
x_{N}(n_{1},n_{2},...,n_{s}),  \label{P}
\end{equation}%
with $n_{1}+n_{2}+n_{3}+...n_{s}=N.$ However, we are interested in the rank
of the observed different values of the macrostates, not in their
distribution of probability. To tackle this problem, we notice that each
value $x_{N}(n_{1},n_{2},...,n_{s})$ corresponds to a \textit{different
macrostate of the system}. In our example, the states $(0,1)$ and $(1,0)$
produce the same global macrostate. These two internal states lead to one
global state that has the same characteristics. If one assume that a certain
characteristic ($X$) of a process or object is a function of $%
n_{1},n_{2},...,n_{s}$, then each value of $X(n_{1},n_{2},...,n_{s})$ can be
mapped to $x_{N}(n_{1},n_{2},...,n_{s})$ and $%
X(n_{1},n_{2},...,n_{s})=X(x_{N}(n_{1},n_{2},...,n_{s}))$. From the previous
considerations, is clear that any rank hierarchy of $%
x_{N}(n_{1},n_{2},...,n_{s})$ will be inherited to $X(n_{1},n_{2},...,n_{s})$%
. Thus, many different rank features of a system are reduced to study the
hierarchy present in $x_{N}(n_{1},n_{2},...,n_{s}).$

For doing such study, there are two cases. In the first, the subsystems are
independent, as in a Bernoulli process, 
\begin{equation}
x_{N}(n_{1},n_{2},...,n_{s})=p_{1}^{n_{1}}p_{2}^{n_{2}}p_{3}^{n_{3}}...p_{s}^{n_{s}},
\label{independent}
\end{equation}%
and the other is the general case of interacting subsystems, in which the
addition of a new subsystem leads to a functional relationship of the type,%
\begin{equation}
x_{N+1}(n_{1},n_{2},...,n_{s})=f(x_{N}(n_{1},n_{2},...,n_{s})).
\end{equation}%
In the next section we will consider only the case of independent
subsystems, in which no extra information is needed in order to model the
correlation in the system. This allows to produce the beta-like function in
a simple form.

$\bigskip $

\section{The rank hierarchy as an algebraic problem$\protect\bigskip $}

$\ \ $For independent subsystems, an inspection of Eq. (\ref{independent})
shows that the rank structure can be reduced to the following algebraic
problem. Take $s$ numbers $p_{1},$ $p_{2},...,p_{s}$ at random
(normalization can be imposed at the end of the process), labeled in such a
way that $p_{1}>p_{2}>...>p_{s}$, and multiply once each number by all the
numbers in the set. With these resulting numbers, repeat the process $N$
times to obtain a set of numbers that have the form $%
p_{1}^{n_{1}}p_{2}^{n_{2}}p_{3}^{n_{3}}...p_{s}^{n_{s}}$, where $%
n_{1}+n_{2}+...+n_{s}=N$. If the resulting numbers are arranged in
decreasing magnitude, we can assign a rank ($r$) to each one according to
its order in the hierarchy. The rank $r=1$ is assigned to $p_{1}^{N}$, while
the lowest rank $r=R$ corresponds to $p_{s}^{N}$. For example, chose at
random three numbers $p_{1}$, $p_{2}$ and $p_{3}$ and form all the possible
products: $p_{1}^{2},p_{1}p_{2},p_{1}p_{3},p_{2}^{2},p_{2}p_{3},p_{3}^{2}$.
In Fig. \ref{numbers}, we present a plot of $\log x_{N}(n_{1},n_{2},n_{3})$
as a function of $r$ for $N=30$ and $p_{1}=0.5202$, $p_{2}=0.3125$ and $%
p_{3}=0.1673$. Fig. \ref{numbers} shows that the resulting ranks are well
fitted by the same two parameter beta-like function, with $a=9.36\pm 0.2$
and $b=10.52\pm 0.2$, with a correlation coefficient of $0.972$. The message
from this numerical experiment is simple: if this product is seen as a
multiplicative process where each number is the probability of making a
certain choice or state in a process, then each possible result has a well
determined hierarchy.

\FRAME{ftbpFU}{3.5163in}{2.6437in}{0pt}{\Qcb{Succesive multiplication of
three numbers $p_{1}=0.5202,p_{2}=0.3125,$ $p_{3}=0.1673$ as a function of
the rank (bold solid line), and a fitting using Eq. (\protect\ref{beta}),
with $a=9.36$, $b=14.53$.}}{\Qlb{numbers}}{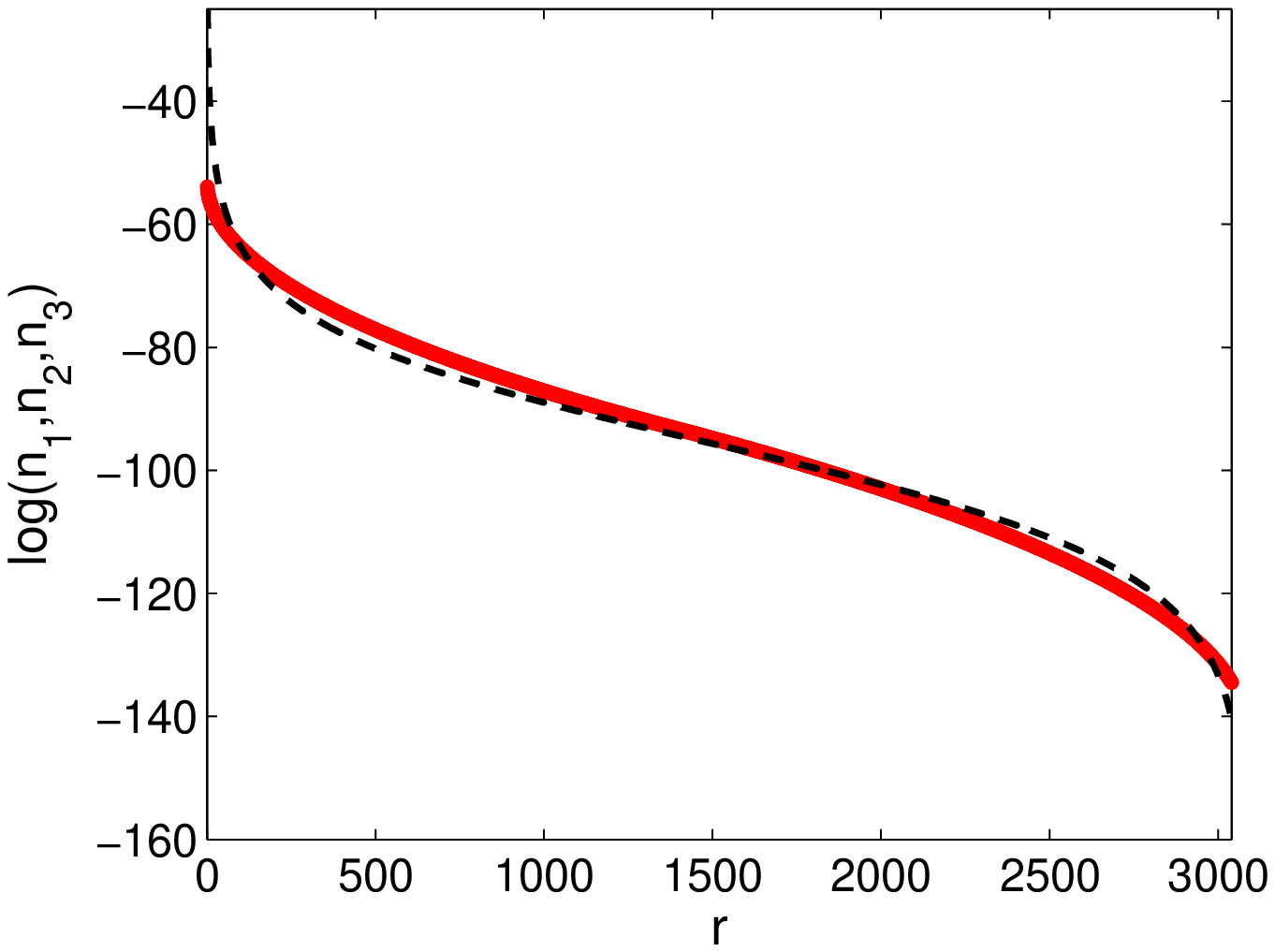}{\special{language
"Scientific Word";type "GRAPHIC";maintain-aspect-ratio TRUE;display
"USEDEF";valid_file "F";width 3.5163in;height 2.6437in;depth
0pt;original-width 5.8219in;original-height 4.3708in;cropleft "0";croptop
"1";cropright "1";cropbottom "0";filename 'numbrank.eps';file-properties
"XNPEU";}}

\bigskip The task that remains is how to calculate $%
x_{N}(n_{1},n_{2},...,n_{s})$ in terms of the rank. The problem is more
easily solved using the logarithm of $x_{N}(n_{1},n_{2},...,n_{s})$,%
\begin{equation}
\log x_{N}(n_{1},n_{2},...,n_{s})=n_{1}\log p_{1}+n_{2}\log
p_{2}+...+n_{s}\log p_{s}.  \label{intlattice}
\end{equation}%
Each set of values $(n_{1},n_{2},...,n_{s})$ is a point with integer
coordinates in a $s-$dimensional space. Since $n_{1}+n_{2}+...+n_{s}=N$, all
the points are in a subspace of dimension $s-1$. The problem of the rank is
reduced to find a path between the maximal rank point (with coordinates $%
\left( N,0,0,...,0\right) $) to the minimum $\left( 0,0,0,...,N\right) $ in
such a way that $\log x_{N}(n_{1},n_{2},...,n_{s})$ decreases in each step.
For $s=2$, the solution is easy to find. Using that $n_{1}+n_{2}=N$, 
\begin{equation}
x_{N}(n_{1},n_{2})=x_{N}(n_{2})=p_{1}^{N-n_{2}}p_{2}^{n_{2}},
\end{equation}%
from where it follows that the range is given by $r=n_{2}+1$. Then,%
\begin{equation}
x_{N}(r)=p_{1}^{N}\left( \frac{p_{2}}{p_{1}}\right)
^{r-1}=p_{1}^{N}e^{-A(r-1)},  \label{seq2}
\end{equation}%
with $A=\left\vert \ln (p_{2}/p_{1})\right\vert $. Eq. (\ref{seq2}) shows
that the numbers decrease in an exponential way as a function of the rank.

The case $s=3$ can be easily visualized in Fig. \ref{triangle}, where the
points in the integer lattice defined by Eq. (\ref{intlattice}) are shown as
circles.

\FRAME{ftbpFU}{5.6317in}{2.936in}{0pt}{\Qcb{Path of decreasing rank in the $%
n_{1},n_{2}$ and $n_{3}$ space, for $N=15$ and three random numbers $%
p_{1}=0.5202,p_{2}=0.3125,$ $p_{3}=0.1673$.}}{\Qlb{triangle}}{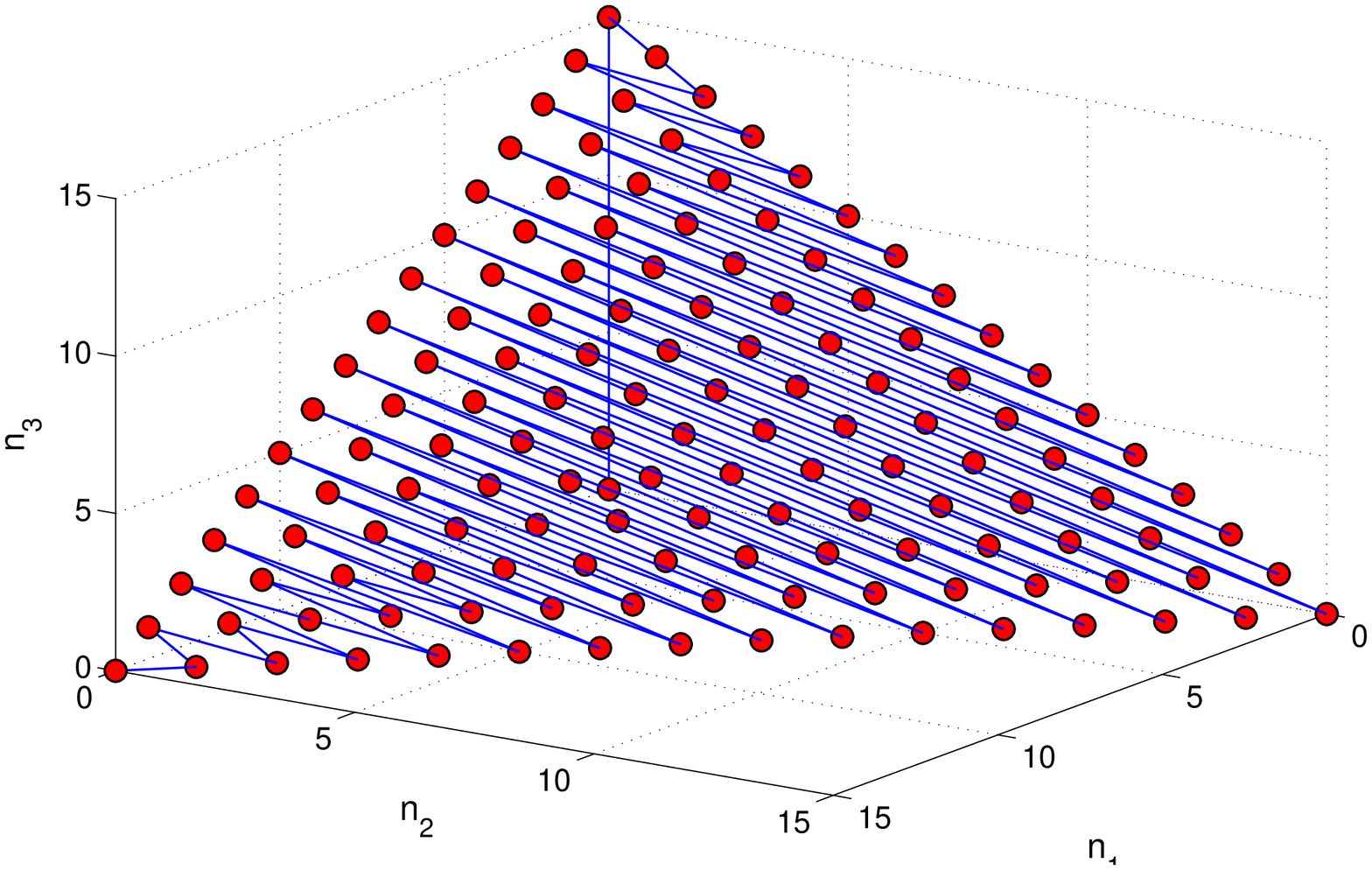}{%
\special{language "Scientific Word";type "GRAPHIC";maintain-aspect-ratio
TRUE;display "USEDEF";valid_file "F";width 5.6317in;height 2.936in;depth
0pt;original-width 10.587in;original-height 5.5011in;cropleft "0";croptop
"1";cropright "1";cropbottom "0";filename 'triangle.eps';file-properties
"XNPEU";}}

A path between points of decreasing $\log x_{N}(n_{1},n_{2},...,n_{s})$ is
indicated as a line that joins the lattice points in Fig. \ref{triangle},
for a given set of numbers $p_{1},p_{2}$ and $p_{3}$. Figure \ref{sequence}
shows how the values of $n_{1},n_{2}$ and $n_{3}$ vary as a function of the
range. A very complicated oscillatory pattern is seen , although a well
defined envelope is also observed. This envelope is in fact the key to solve
the problem, since it is the responsible of the ranking behavior. Notice
also that all paths always start at $(N,0,0)$ and finish at $(0,0,N)$, since 
$\log p_{1}>\log p_{2}>\log p_{3}.$ \FRAME{ftbpFU}{4.1848in}{2.4137in}{0pt}{%
\Qcb{Values of $n_{1}$ (thin solid line), $n_{2}$ (grey line) and $n_{3}$
(solid bold line) as a function of the rank, for $N=20$ and $%
p_{1}=0.5202,p_{2}=0.3125,$ $p_{3}=0.1673$.}}{\Qlb{sequence}}{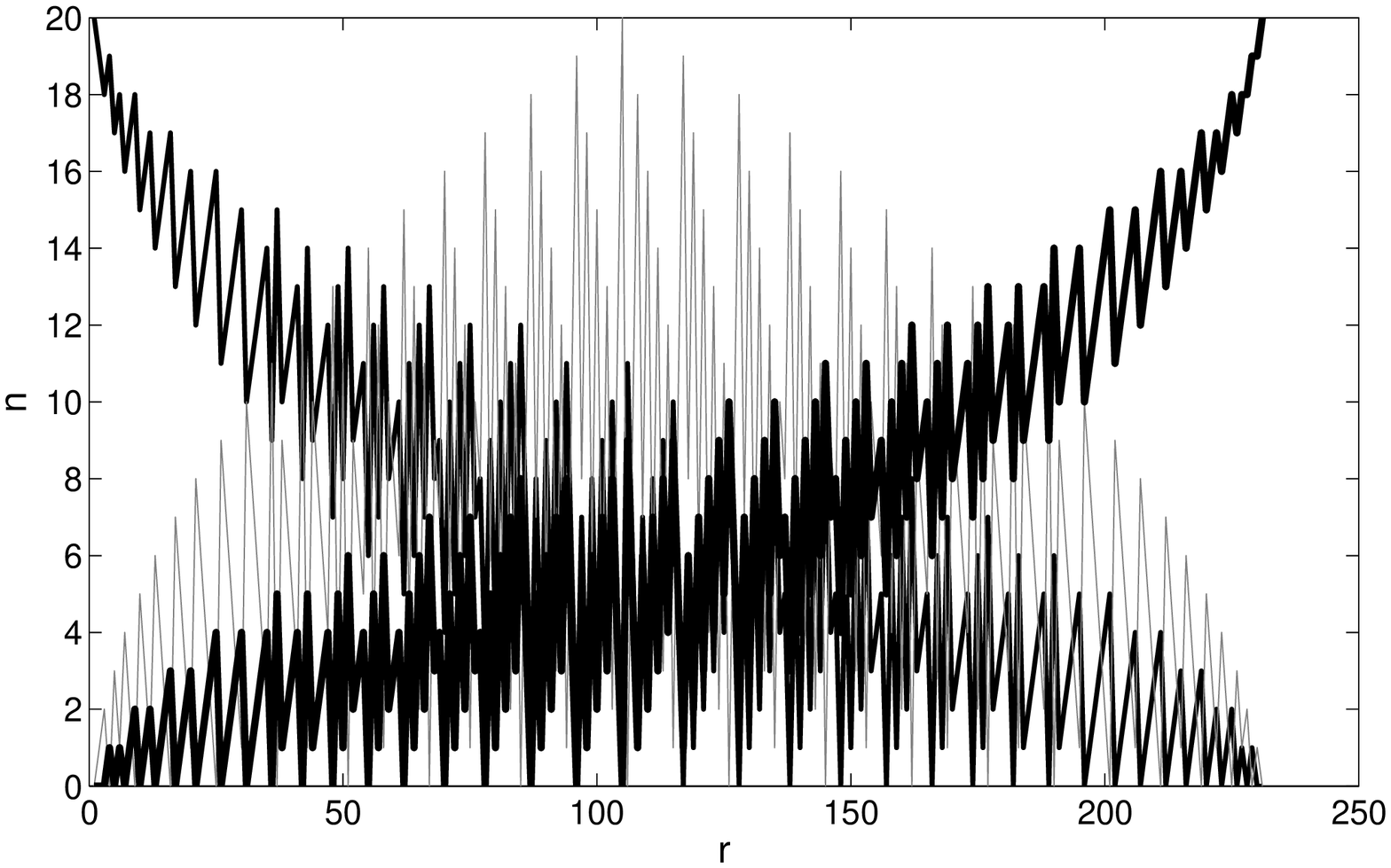}{%
\special{language "Scientific Word";type "GRAPHIC";maintain-aspect-ratio
TRUE;display "USEDEF";valid_file "F";width 4.1848in;height 2.4137in;depth
0pt;original-width 9.5415in;original-height 5.489in;cropleft "0";croptop
"1";cropright "1";cropbottom "0";filename 'path.eps';file-properties
"XNPEU";}}

In general, since the index $n_{j}$ is a function of the rank $r$, \ we can
write that $n_{j}=n_{j}(r)$ where $r$ is just the number of steps used to go
from the point $(N,0,...,0)$ to a certain $(n_{1},n_{2},n_{3},...,n_{s})$.
It follows that,%
\begin{equation}
\log x_{N}(r)=n_{1}(r)\log p_{1}+n_{2}(r)\log p_{2}+...+n_{s}(r)\log p_{s}
\end{equation}%
The task is reduced to find the functions $n_{j}(r)$ for a given set $%
\{p_{j}\}.$ Consider again the case of an initial set of three numbers, $s=3$%
. Using that $n_{1}+n_{2}+n_{3}=N$, $\log x_{N}(r)$ can be written as,

\begin{equation}
\log x_{N}(r)=N\log p_{1}+n_{2}(r)\log \delta _{21}+n_{3}(r)\log \delta
_{31}.  \label{2D}
\end{equation}%
with $\delta _{21}=p_{2}/p_{1}$ and $\delta _{31}=p_{3}/p_{1}.$ The solution
for any set $p_{1},p_{2}$ $,p_{3}$ is complicated, because some paths are
not periodic. However, one can work out first the cases $p_{1}\sim p_{2}$ $%
\gg p_{3}$ and $p_{1}\gg p_{2}\sim $ $p_{3}$ that give insights about how to
treat others.

\FRAME{ftbpFU}{3.9539in}{2.9732in}{0pt}{\Qcb{Path of decreasing ranks in the 
$n_{2}$ and $n_{3}$ plane for $p_{1}\sim p_{2}$ $\gg p_{3}$, where the $%
n_{1} $ coordinate was eliminated using that $n_{1}+n_{2}+n_{3}=N$. The
dotted line corresponds to all the $n_{2MAX}(r)$, which defines the envelope
of the ranking sequence.}}{\Qlb{triplane}}{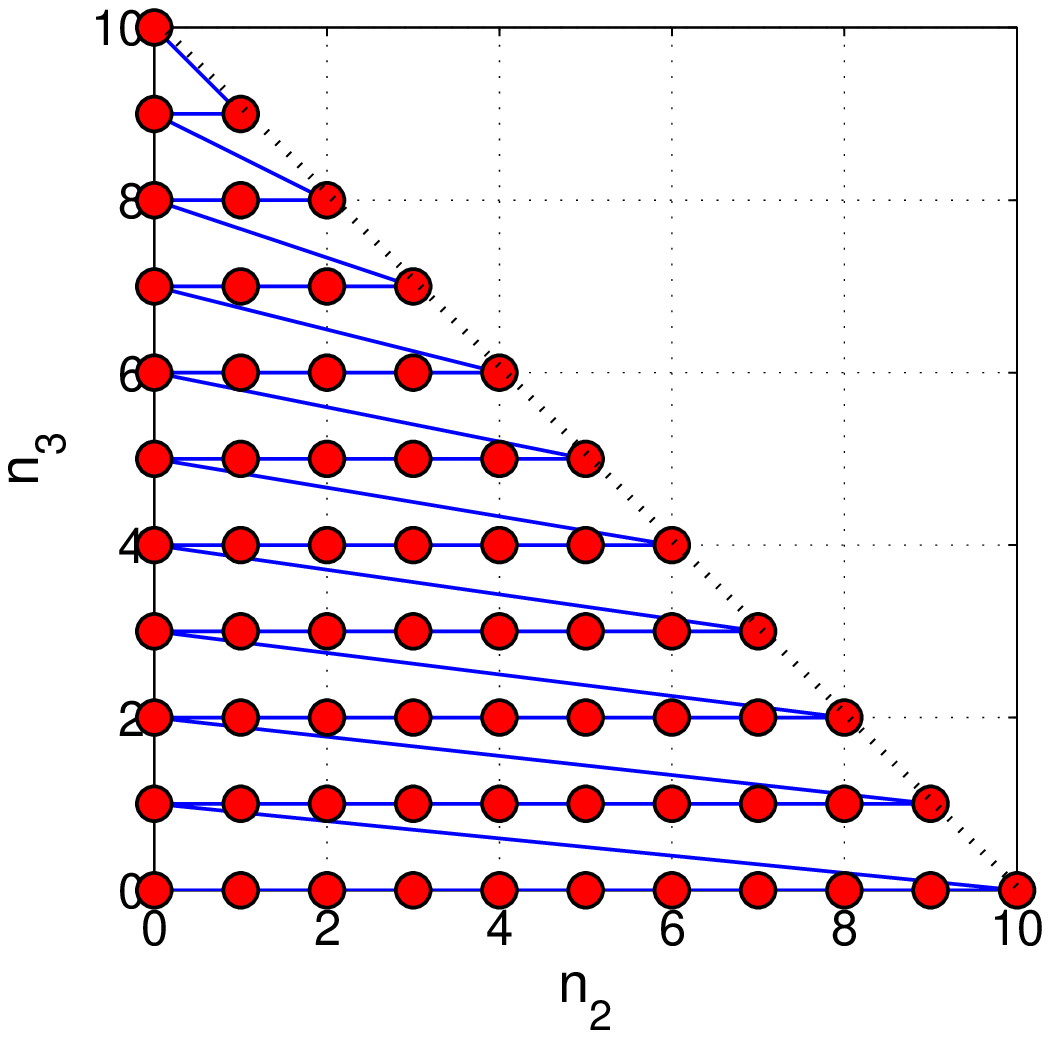}{\special{language
"Scientific Word";type "GRAPHIC";maintain-aspect-ratio TRUE;display
"USEDEF";valid_file "F";width 3.9539in;height 2.9732in;depth
0pt;original-width 5.8219in;original-height 4.3708in;cropleft "0";croptop
"1";cropright "1";cropbottom "0";filename 'triplane.eps';file-properties
"XNPEU";}}

Let us first consider the limit $p_{1}\sim p_{2}$ $\gg p_{3},$ and $\delta
_{21}^{2}\gg \delta _{31}$. The corresponding path is easy to find because
it is similar to an odometer with an increased range after each turn, as
seen in Fig. \ref{triplane}, due to the hierarchy $1>\delta _{21}>\delta
_{21}^{2}>\delta _{31}>\delta _{21}\delta _{31}>\delta _{31}^{2}>...>\delta
_{31}^{N}$. For example, when $N=2$ this leads to the following table that
contains the number $x_{N}(r)$ as a function of the rank, and the
corresponding path given by $n_{2}$ and $n_{3}$,

\begin{center}
\begin{tabular}{|l|l|l|l|l|}
\hline
$x_{N}(r)$ & $n_{2}$ & $n_{3}$ & $r$ & $n_{2M}(r)$ \\ \hline\hline
$p_{1}^{2}$ & $0$ & $0$ & $1$ & $-$ \\ \hline
$p_{1}^{2}\delta _{21}$ & $1$ & $0$ & $2$ & $-$ \\ \hline
$p_{1}^{2}\delta _{21}^{2}$ & $2$ & $0$ & $3$ & $2$ \\ \hline
$p_{1}^{2}\delta _{31}$ & $0$ & $1$ & $4$ & $-$ \\ \hline
$p_{1}^{2}\delta _{21}\delta _{31}$ & $1$ & $1$ & $5$ & $1$ \\ \hline
$p_{1}^{2}\delta _{31}^{2}$ & $0$ & $2$ & $6$ & $0$ \\ \hline
\end{tabular}
\end{center}

The sequence of the path goes as follows, first $n_{2}(r)$ is increased one
by one as $n_{3}$ remains constant, until \ it reaches a maximal value
called $n_{2MAX}(r)$ which in fact determines the envelope of the ranking
sequence and thus the basic shape of the curve $x_{N}(r)$ (the envelope that
contains $n_{2MAX}(r)$ is shown in Fig. \ref{triplane} as a dotted line).
Once $n_{2}(r)$ increases from zero to $n_{2MAX}(r)$, a new cycle begins
with $n_{2}(r)=0$ and $n_{3}(r+1)=n_{3}(r)+1.$ As a result, the number of
steps $r$ to reach $n_{2MAX}(r)$ is given by,%
\begin{equation}
R-r\approx n_{2MAX}(r)+\sum_{j=1}^{n_{2MAX}(r)}j=n_{2MAX}(r)+\frac{%
n_{2MAX}(r)\left( n_{2MAX}(r)+1\right) }{2},  \label{n2max}
\end{equation}%
where $R$ is the maximal rank. Then,%
\begin{equation}
n_{2MAX}(r)\approx N\left( 1-\frac{r}{R}\right) ^{1/2}
\end{equation}%
The corresponding value of $n_{3}(r)$ can be obtained from the condition $%
n_{2}+n_{3}\leq N$. Finally, the number as a function of the rank is given
by,%
\begin{equation}
x_{N}(r)\approx \left[ p_{1}\left( \frac{p_{2}}{p_{1}}\right) ^{\left( 1-%
\frac{r}{R}\right) ^{1/2}}\left( \frac{p_{3}}{p_{1}}\right) ^{1-\left( 1-%
\frac{r}{R}\right) ^{1/2}}\right] ^{N}.  \label{case1}
\end{equation}%
Figure \ref{limit1} shows the excellent agreement between Eq. (\ref{case1})
and the curve obtained for $p_{1}=0.5250$, $p_{2}=0.4250$, $p_{3}=0.000047$.
Furthermore, Eq. (\ref{case1}) can be written as an stretched exponential as
follows, 
\begin{equation}
x_{N}(r)\approx p_{3}^{N}\exp \left[ D\left( 1-\frac{r}{R}\right) ^{1/2}%
\right] ,  \label{stret1}
\end{equation}%
with $D=N\left\vert \log (p_{2}/p_{3})\right\vert $ and $R$ is the maximal
value of $r$. Notice in Fig. \ref{limit1} how this formula works better as
the rank approaches $R.$

\bigskip \FRAME{ftbpFU}{3.2872in}{2.4725in}{0pt}{\Qcb{Numerical results for
the ranking of the succesive product of three numbers such that $p_{1}\sim
p_{2}$ $\gg p_{3}$. The smooth line is the prediction using Eq. (\protect\ref%
{case1}). }}{\Qlb{limit1}}{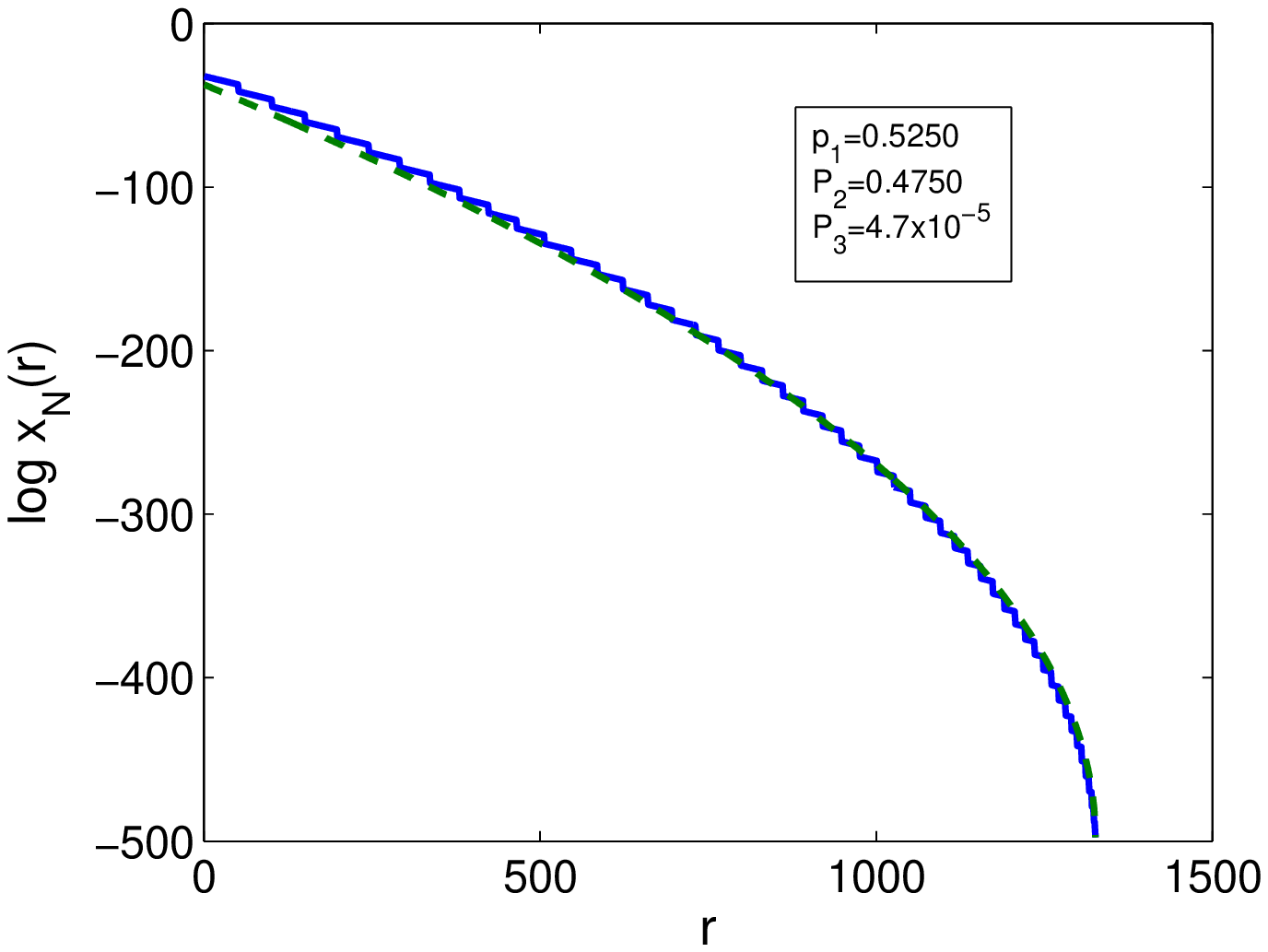}{\special{language "Scientific
Word";type "GRAPHIC";maintain-aspect-ratio TRUE;display "USEDEF";valid_file
"F";width 3.2872in;height 2.4725in;depth 0pt;original-width
5.8219in;original-height 4.3708in;cropleft "0";croptop "1";cropright
"1";cropbottom "0";filename 'limite1.eps';file-properties "XNPEU";}}

The case $p_{1}\gg p_{2}\sim $ $p_{3}$ can be tackled in a similar way. The
result is,%
\begin{equation}
x_{N}(r)\approx p_{1}^{N}\exp \left[ -E\left( \frac{r}{R}\right) ^{1/2}%
\right] .  \label{case2}
\end{equation}%
with $E=N\left( \log (p_{1}/p_{3})-\log (p_{2}/p_{3})\right) $, and as shown
in Fig. \ref{limit2}, the agreement is also good, specially for low values
of $r$.

\FRAME{ftbpFU}{2.8184in}{2.1197in}{0pt}{\Qcb{Ranking of the succesive
product of three numbers such that $p_{1}\gg p_{2}\sim $ $p_{3}$, for $%
p_{1}=0.99999$, $p_{2}=6.2\times 10^{-6}$, $p_{3}=3.8\times 10^{-6}$. The
dashed line is the prediction made from Eq. (\protect\ref{case2}), compared
with the numerical result for $N=100$ iterations (solid line). }}{\Qlb{limit2%
}}{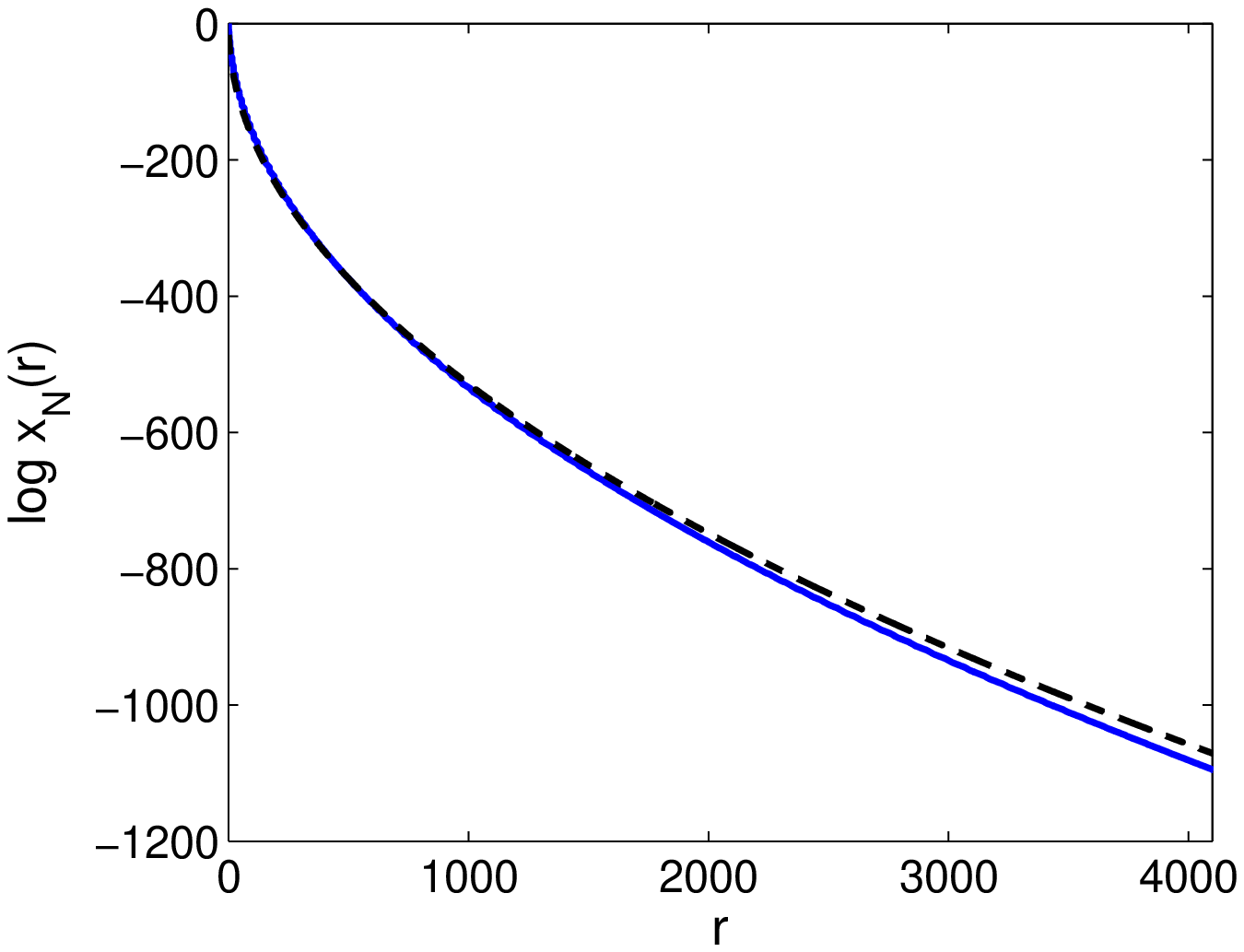}{\special{language "Scientific Word";type
"GRAPHIC";maintain-aspect-ratio TRUE;display "USEDEF";valid_file "F";width
2.8184in;height 2.1197in;depth 0pt;original-width 5.8219in;original-height
4.3708in;cropleft "0";croptop "1";cropright "1";cropbottom "0";filename
'limite2.eps';file-properties "XNPEU";}}

Now consider the general case in which $p_{1,}p_{2}$ and $p_{3}$ have the
same order of magnitude, as in Fig. \ref{numbers}, where two tails appears,
one for small $r$ and the other at $r$ near $R$. The tail at low $r$ is
produced basically by the hierarchy in the biggest probabilities, \textit{%
i.e.}, by numbers where $n_{1}\sim N$. In a similar way, the tail for $r$
near $R$ is produced by the lowest probability hierarchy, $n_{3}\sim N$.
These dominant factors are due to large statistical deviations and are the
origin of the long tails in the otherwise power law observed in the ranks.
The main effect in these tails when $p\approx p_{2}\approx p_{3}$ is that
the sequence of ordering is not uniform as can be observed in Fig. \ref%
{numbers}, for which a very complicate path appears. As a result, Eq. (\ref%
{n2max}) changes with the apparition of new subcycles in the rank path.
These changes are the result of the increasing number of cycles in the
odometer that we have discussed, as is also clear from the change in the
exponents that are transformed from $1$ to $1/2$ as $s$ goes from $s=2$ to $%
s=3$. Eq. (\ref{stret1}) is thus transformed into a generalized expression,%
\begin{equation}
x_{N}(r)\approx p_{3}^{N}\exp \left[ D\left( 1-\frac{r}{R}\right) ^{\beta }%
\right]  \label{limiting1}
\end{equation}%
in which $\beta $ is a yet unknown exponent, always less than one. In a
similar way, Eq. (\ref{case2}) \ should be replaced by, 
\begin{equation}
x_{N}(r)\approx p_{1}^{N}\exp \left[ -E\left( \frac{r}{R}\right) ^{\alpha }%
\right] .  \label{limiting2}
\end{equation}%
with $\alpha <1$. These generic exponents for the tails also appear for $s>3$
since from the polynomial equivalent to Eq. (\ref{n2max}), one gets $\alpha
\approx \beta \approx 1/(s-1)$. A simple procedure to combine the tails
represented by Eq. (\ref{limiting1}) and Eq. (\ref{limiting2}) is obtained
by making the observation that for a given tail, only one stretched
exponential produces a curved tails in a semi-log plot, while the other
tends toward a constant, \textit{i.e.}, if we consider the derivative of Eq.
(\ref{limiting1}),

\begin{equation}
\left( \frac{d\ln x_{N}(r)}{dr}\right) =-\frac{\beta D}{R}\left( 1-\frac{r}{R%
}\right) ^{\beta -1}  \label{derivative}
\end{equation}%
is clear that $x_{N}^{\prime }(r)$ is nearly a constant if $r\ll 1$,
corresponding to the limit in which Eq. (\ref{limiting2}) has greater
curvature. Analyzing the limit $r\rightarrow R$ gives a similar result,%
\begin{equation}
\left( \frac{d\ln x_{N}(r)}{dr}\right) =-\frac{\alpha E}{R}\left( \frac{r}{R}%
\right) ^{\alpha -1}.  \label{derivative2}
\end{equation}

From these considerations, a simple way to produce a function with the
required dependences when $r\rightarrow R$ and $r\rightarrow 1$ is the
following,

\begin{equation}
x_{N}(r)\approx C_{1}\exp \left[ D\left( 1-\frac{r-1}{R}\right) ^{\beta }%
\right] \exp \left[ -E\left( \frac{r}{R}\right) ^{\alpha }\right] ,
\label{final}
\end{equation}%
where $C_{1}$ is a constant. A plot of the previous expression is presented
in Fig. \ref{generalplot}, showing the basic shape of the studied beta
function.

\FRAME{ftbpFU}{3.1522in}{2.3696in}{0pt}{\Qcb{A plot of Eq. (\protect\ref%
{final}) using $C_{1}=2$, $D=1$, $E=1$, $\protect\beta =1$ and $\protect%
\alpha =1$.}}{\Qlb{generalplot}}{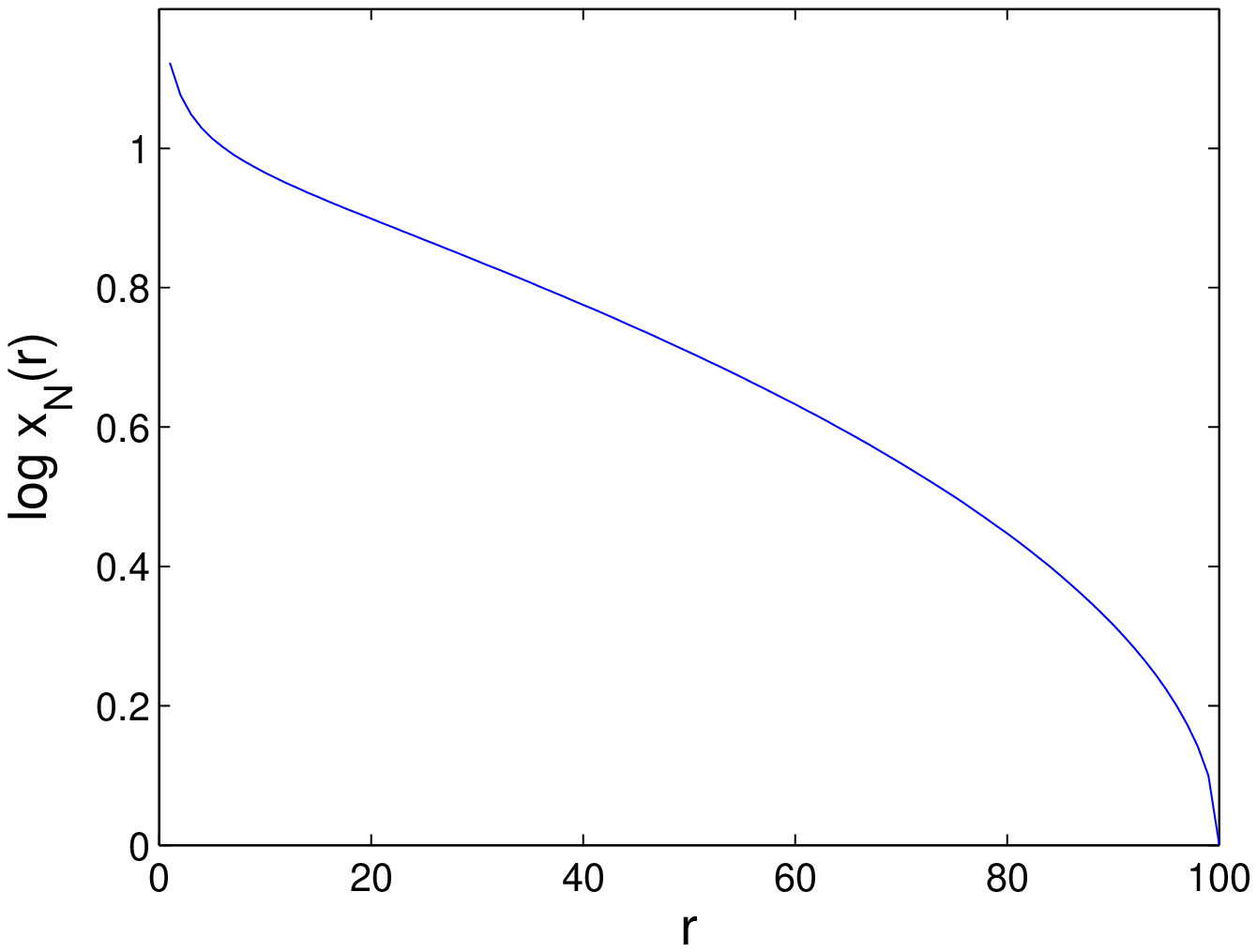}{\special{language
"Scientific Word";type "GRAPHIC";maintain-aspect-ratio TRUE;display
"USEDEF";valid_file "F";width 3.1522in;height 2.3696in;depth
0pt;original-width 5.8219in;original-height 4.3708in;cropleft "0";croptop
"1";cropright "1";cropbottom "0";filename 'generalstret.eps';file-properties
"XNPEU";}}

Finally, Eq. (\ref{final}) can be simplified when many states are present
since for $s\gg 1$, $\alpha \approx \beta \approx 1/(s-1)$ and thus $\alpha
\rightarrow 0$ and $\beta \rightarrow 0.$ Then, by using the observation
about the derivatives that appears in Eq. (\ref{derivative}) and Eq. (\ref%
{derivative2}), one can approximate the derivatives like in Eq. (\ref%
{derivative}) as follows, 
\begin{equation}
\left( \frac{d\log x_{N}(r)}{dr}\right) =-\frac{\beta D}{R}\left( 1-\frac{r}{%
R}\right) ^{\beta -1}\approx -\frac{\beta D}{R}\left( 1-\frac{r}{R}\right)
^{-1}.
\end{equation}%
A similar thing can be done in the tail $r\rightarrow 1$, for which $\alpha $
can be neglected with respect to one in Eq. (\ref{derivative2}). Combining
both tails in a sole expression we get,%
\[
\left( \frac{d\log x_{N}(r)}{dr}\right) \approx -\frac{\beta D}{R}\left( 1-%
\frac{r}{R}\right) ^{-1}-\frac{\alpha E}{R}\left( \frac{r}{R}\right) ^{-1}. 
\]%
By integrating the previous equation, we finally obtain the beta-like
function given by Eq. (\ref{beta}), where the exponents $a$ and $b$ are
given by,%
\begin{equation}
a=\alpha E\text{ \ and \ }b=\beta D
\end{equation}%
Thus, the beta-like function is obtained when we have a large number of
states in the system. Notice how the parameters $a$ and $b$ are determined
mainly by the behavior in the tails.

\section{Conclusions \ }

In conclusion, we found a simple formula that allows to fit many different
rank phenomena. This formula shows that there is a certain universality at
the tails, explained by considering the ranking of a multiplicative process.
We have shown that such problem is equivalent to an algebraic problem: find
the rank of the successive product of numbers. A task that remains is to how
to get the coefficients $a$ and $b$ from physical principles, using for
example master equations and the concept of multiscaling modelling. A key
observation for such study is that for expansion-modification algorithms in
DNA models, $a>b$ if the expansion probability of the genetic code is bigger
the than mutation rate \cite{Mansilla}. Thus, $a$ and $b$ represent the
relative influence of two general mechanisms, where each of them dominate at
a given tail. According to some preliminary results, $a$ seems to be related
with a certain funnel type of energy landscape, as in protein folding, which
leads to a certain deterministic sequence, while $b$ is associated with a
many valley landscape, as seen in spin glasses. This last opposite effect
provides much more variability in the sequence of results. Such correlation
is consistent with associating $b$ to the stochastic component of the
dynamics and $a$ with the most deterministic features \cite{Mansilla}. In
future work, we will elucidate with more detail such mechanisms.

Acknowledgments. This work was supported by DGAPA-UNAM project IN-117806,
CONACyT 48783-F and 50368.


\bigskip

\begin{thebibliography}{99}
\bibitem{Li} Li W., Phys. Rev. E \textbf{43}, 5240 (1991), see also: Li W.,
http://www.nslij-genetics.org/wli/zipf/\ (2003).

\bibitem{Bretz} M. Bretz, R. Zaretzki, S.B. Field, N. Mitarai and F. Nori,
Europhysics Lett. \textbf{74} (2006) 1116 .

\bibitem{Nuclear} G. Audi, O. Bersillon, J. Blachot and A. H. Wapstra,
Nuclear Physics \textbf{A624} (1997) 124.

\bibitem{Fortunato} S. Fortunato, A. Flammini, and F. Menczer, Phys. Rev.
Lett. \textbf{96} (2006) 218701.

\bibitem{Yang} A.C.C. Yang, S.S. Hseu, H.W Yien, A.L. Goldberger, and C.K.
Peng, Phys. Rev. Lett. \textbf{90} (2003) 108103.

\bibitem{Jeong} H. Jeong, B. Tombor, R. Albert, Z.N. Oltvai, A.L. Barabasi,
Nature \textbf{407} (2000) 651.

\bibitem{Lequan} Le Quan H., Sicilia-Garc\'{\i}a E.I., Minj J. and Smith
F.J., \textit{\ Proceedings of the 17th. International Conference on
Computer Lingustics, }Montreal, (2002).

\bibitem{Sornette} J. Laherrere and D. Sornette, Eur. Phys. J.B. \textbf{2}
(1998) 525 .

\bibitem{Montroll} E.W. Montroll and M.F. Shlesinger, J. of Statistical
Physics \textbf{32} (1983) 209.

\bibitem{Kolmogorov1} A.N. Kolmogorov, Dokl. Akad. Nauk SSSR \textbf{30}
(1941) 299 (reprinted in Proc. R. Soc. Lond. A \textbf{434} (1991) 9 ).

\bibitem{Kolmogorov2} A.N. Kolmogorov, Dokl. Akad. Nauk SSSR \textbf{32}
(1941) 16 (reprinted in Proc. R. Soc. Lond. A \textbf{434} (1991) 15).

\bibitem{Kolmogorov3} A. Kolmogorov, J. Fluid. Mech. \textbf{13} (1962) 82 .

\bibitem{Warhaft} Z. Warhaft, Annu. Rev. Fluid Mech. \textbf{32}, (2000) 203.

\bibitem{Moyano} L.G. Moyano, C. Tsallis and M. Gell-Mann, Europhys. Lett. 
\textbf{72} (2006) 355.

\bibitem{Marsh} J. A. Marsh, M.A. Fuentes, L.G. Moyano and C. Tsallis,
Physica A \textbf{372} (2006) 183 .

\bibitem{Pospescu} Popescu I., Glottometrics \textbf{6} (2003) 83.

\bibitem{database} Codon Usage Database, NCBI-GenBank
http://www.kazuza.or.jp/codon

\bibitem{Cocho} G. Cocho, G. Mart\'{\i}nez-Mekler, to be published.

\bibitem{Manrubia} S.C. Manrubia, D.H. Zanette, Phys. Rev. E \textbf{59}
(1999) 4945 .

\bibitem{Naumis} G. Naumis, G. Cocho, to be published.

\bibitem{Mansilla} Mansilla R, Cocho G., Complex Systems. \textbf{12} (2000)
207.
\end{thebibliography}
\end{document}